\documentstyle[12pt]{article}

\begin{document}
\baselineskip=18pt
\begin{titlepage}
\begin{center}
\large{\bf BILINEARIZATION OF $N=1$ SUPERSYMMETRIC MODIFIED
KDV EQUATIONS}
\end{center}
\vspace{2 cm}
\begin{center}
{\bf Sasanka Ghosh}\footnote{email: sasanka@iitg.ernet.in} and
{\bf Debojit Sarma}\footnote{email: debojit@iitg.ernet.in}\\
{\it Department of Physics, Indian Institute of Technology,}\\
{\it North Guwahati, Guwahati 781039, INDIA}\\
\end{center}
\vspace{1 cm}


\vspace{3 cm}
\begin{abstract}
Two different types of $N=1$ modified KdV equations are shown to possess 
$N$ soliton solutions. The soliton solutions of these equations are obtained 
by casting the equations in the bilinear forms using the supersymmetric 
extension of the Hirota method. The distinguishing features of the soliton 
solutions of $N=1$ mKdV and $N=1$ mKdV B equations are discussed. 
\end{abstract}
\end{titlepage}

\newpage

\section{Introduction}

The study of supersymmetric integrable systems has assumed great importance 
in recent years. This has been motivated by a number of factors, one of them
being the possibility that a fundamental physical theory must be supersymmetric.
In the context of integrable models, this has resulted in the supersymmetrization
of a number of integrable equations and the extension of the methodologies
involved in the study of integrable hierarchies to the supersymmetric 
framework. Investigations have also been made in recent times to obtain 
$\tau$ functions for supersymmetric integrable hierarchies as they are 
essential in the context of integrability as well as in generating soliton 
solutions \cite{Aratyn,Yung,Carstea,Ghosh}. Integrability of a nonlinear
differential equation implies the existence of nondissipating wave like solutions
called solitons. Of the various known methods to generate soliton solutions,
the Hirota bilinear method is the most direct and elegant one 
\cite{Hirota}. $\tau$ functions play the central role in obtaining soliton 
solutions in the Hirota formalism. Another significance of the $\tau$ 
function is its role as the effective action in quantum field theories.

Extension  of the bilinear formalism to supersymmetric systems has been a
comparatively recent development. A number of supersymmetric systems
in $N=1$ superspace have been bilinearized \cite{Yung,Carstea} and their 
soliton solutions constructed via this technique. For $N=2$ integrable systems, 
in a recent work, the $N$ soliton solutions of the supersymmetric KdV of 
Inami and Kanno \cite{Inami} have found through the super Hirota formalism
\cite{Ghosh}. However, several $N=1$ equations still remain which have not 
been bilinearized to obtain $N$ soliton solutions. $N=1$ modified KdV and 
modified KP hierarchies are such important examples, related to the 
supersymmetric KdV and KP hierarchies.

In this paper we consider $N=1$ mKdV equation \cite{MBN} involving spin $1/2$
super field $\psi (x,\theta,t)$,
\begin{equation}
\partial_{t}\psi+D^{6}\psi-3\psi D^{3}\psi D\psi-3(D\psi)^{2}D^{2}\psi=0
\label{1}
\end{equation}
and $N=1$ mKdV B equation having spin $1$ super field $\Psi (x,\theta,t)$
\begin{equation}
\partial_{t}\Psi+D^{6}\Psi-2D^{2}\Psi^{3}=0
\label{2}
\end{equation}
and show that these equations can be bilinearised following the Hirota method and
possess $N$ soliton solutions. In (\ref{1},\ref{2}) $D$ denotes the
superderivative defined by
\begin{equation}
D=\frac{\partial}{\partial\theta}+\theta\frac{\partial}{\partial x}
\label{3}
\end{equation}
While the bilinear forms of the first system (\ref{1}) involve the super 
Hirota operator \cite{Yung}, it will be seen that the for the second system 
(\ref{2}) will involve only the bosonic or ordinary Hirota operator in its 
bilinear equations. But the importance of $N=1$ mKdV B vis a vis
$N=1$ KdV B arises because of their connections with superstring theory from
the point of view of matrix models \cite{Becker}. The consequence of the presence
or absence of the super Hirota operator in the  bilinear forms of 
these two systems will be reflected in the 
soliton solutions of the equations in the role played by the fermionic 
parameters and it will also be shown that they possess distinctly different soliton 
solutions. These two equations, in fact, have different origins as the
reductions of $N=2$ systems.

The organization of the paper is as follows. In sections 1 and 2, the
bilinearization of the $N=1$ KdV and $N=1$ mKdV B equations respectively will 
be demonstrated. Section 3 will be a discussion of the one soliton
solution for both equations. In section 4, the existence of two and higher
solitons will be shown. Section 5 is the concluding one.

\setcounter{equation}{0}

\section{$N=1$ mKdV equation}

The $N=1$ mKdV equation (\ref{1}) is related to the $N=1$ KdV equation of
Manin-Radul-Mathieu \cite{Manin,Mathieu}
\begin{equation}
\partial_{t}\phi+D^{6}\phi+3D^{2}(\phi D\phi)=0
\label{1a}
\end{equation}
through the super Miura transformation \cite{Mathieu1}
\begin{equation}
\phi=D^{2}\psi+\psi D\psi
\label{1b}
\end{equation}
Here the superfields $\psi$ and $\phi$ are both fermionic having spins $3/2$ and
$1/2$ respectively. The $N=1$ mKdV also follows as an $N=1$ reduction of the
$N=2$ mKdV system \cite{Inami}, which has the explicit form
\begin{equation}
\partial_{t}\psi_{1}=-D\left [D^{5}\psi_{1}+3\psi_{1}D^{2}\psi_{2}D\psi_{2}
-\frac{1}{2}(D\psi_{1})^{3}-\frac{3}{2}D\psi_{1}(D\psi_{2})^{2}\right]
\label{1c}
\end{equation}
\begin{equation}
\partial_{t}\psi_{2}=-D\left [D^{5}\psi_{2}+3\psi_{2}D^{2}\psi_{1}D\psi_{1}
-\frac{1}{2}(D\psi_{2})^{3}-\frac{3}{2}D\psi_{2}(D\psi_{1})^{2}\right]
\label{1d}
\end{equation}
By imposing the constraint $\psi_{1}=-\psi_{2}=\psi$, (\ref{1c},\ref{1d})
immediately reduce
to (\ref{1}). But we will see that (\ref{2}) is not a direct reduction of
$N=2$ mKdV. It has a different origin, rather follows from $N=2$ KdV of the
Inami Kanno type.

In order to cast the $N=1$ mKdV equation in the bilinear form the superfield
$\psi$ is written in terms of $\tau$ functions as
\begin{equation}
\psi=D\log\frac{\tau_{1}}{\tau_{2}}
\label{1e}
\end{equation}
In terms of the supersymmetric Hirota derivative $\bf{S}$ \cite{Yung},
defined by
\begin{equation}
{\bf{S}\bf{D}}^{n}_{x}f.g=(D_{\theta_{1}}-D_{\theta_{2}})(\partial_{x_{1}}-
\partial_{x_{2}})^{n}f(x_{1},\theta_{1})g(x_{2},\theta_{2})|_{\begin{array}{c} x_{1}=x_{2}=x \\
\theta_{1}=\theta_{2}=\theta\end{array}}
\label{1f}
\end{equation}
where $\bf{D}$ is ordinary (or bosonic) Hirota derivative, the $N=1$
mKdV equations can be cast in the following bilinear forms:
\begin{equation}
({\bf{S}}{\bf{D}}_{t}+{\bf{S}}{\bf{D}}^{3}_{x})(\tau_{1}.\tau_{1})=0
\label{1g}
\end{equation}
\begin{equation}
({\bf{S}}{\bf{D}}_{t}+{\bf{S}}{\bf{D}}^{3}_{x})(\tau_{2}.\tau_{2})=0
\label{1h}
\end{equation}
\begin{equation}
{\bf{D}}^{2}_{x}(\tau_{1}.\tau_{2})=0
\label{1i}
\end{equation}
\begin{equation}
{\bf{S}}{\bf{D}}_{x}(\tau_{1}.\tau_{2})=0
\label{1j}
\end{equation}
In obtaining the bilinear equations above, use has been made of the identities
\begin{equation}
D^{4}\log (\tau_{1}\tau_{2})+\left (D^{2}\log\frac{\tau_{1}}{\tau_{2}}\right )^{2}
=\frac{{\bf D}^{2}(\tau_{1}.\tau_{2})}{\tau_{1}\tau_{2}}
\label{1k}
\end{equation}
and
\begin{equation}
D^{3}\log (\tau_{1}\tau_{2})+D^{2}\log\frac{\tau_{1}}{\tau_{2}}D\log\frac{\tau_{1}}{\tau_{2}}
=\frac{{\bf S}{\bf D}(\tau_{1}.\tau_{2})}{\tau_{1}\tau_{2}}
\label{1l}
\end{equation}
The bilinear forms (\ref{1g},\ref{1h},\ref{1i},\ref{1j}) involve
supersymmetric operator ${\bf S}$ in addition to the bosonic Hirota
derivative ${\bf D}$. The consequence of the presence of ${\bf S}$ operator
will be apparent in the higher soliton solutions through the
nontrivial relations between bosonic and fermionic parameters.

\setcounter{equation}{0}

\section{$N=1$ mKdV B equation}

Just as the $N=1$ mKdV equation (\ref{1}) is related to the $N=1$ KdV equation of
Manin-Radul-Mathieu through super Miura transformation (\ref{1b}), the $N=1$
mKdV B equation (\ref{2}) is related to the $N=1$ KdV B equation \cite{Becker,Jose}
\begin{equation}
\partial_{t}\Phi+D^{6}\Phi+6D^{2}\Phi D\Phi=0
\label{2a}
\end{equation}
$\Phi(x,\theta)$ being a spin $3/2$ superfield, through a Miura tranformation,
\begin{equation}
D\Phi=D^{2}\Psi-\Psi^{2}
\label{2b}
\end{equation}
Note that the Miura transformation (\ref{2b}) is nonlocal. Interestingly,
in the bosonic limit (\ref{2}) also reduces to the modified KdV equation
and is invariant under the supersymmetric transformation
\begin{equation}
\delta\Psi^{b}=\eta\Psi^{f}\;\; ; \;\; \delta\Psi^{f}=\eta\partial_{x}\Psi^{b}
\label{2c}
\end{equation}
As mentioned earlier, the equation (\ref{2}) directly results from $N=2$
KdV equations \cite{Inami}
\begin{equation}
\partial_{t}U+D^{6}U+3D^{2}((DU)V)-\frac{1}{2}D^{2}(U^{3})=0
\label{2d}
\end{equation}
\begin{equation}
\partial_{t}V+D^{6}V-3D^{2}(V(DV))-\frac{3}{2}D^{2}(VU^{2})+3D^{2}(VD^{2}U)=0
\label{2e}
\end{equation}
where $U$ and $V$ are superfields of conformal spin $1$ and $3/2$
respectively. It is straightforward to show that in the limit $V=0$ reduces
to the $N=1$ mKdV B equation by identifying $U=\Psi$. Interestingly, in yet
another limit, namely $V=DU$ both the equations (\ref{2d},\ref{2e}) acquire 
the same form which is identical to the $N=1$ mKdV B equation.

With the substitution of
\begin{equation}
\Psi=D^{2}\log\frac{\tau_{1}}{\tau_{2}}
\label{2f}
\end{equation}
the $N=1$ mKdV B can be cast in the bilinear equations
\begin{equation}
({\bf{D}}_{x}{\bf{D}}_{t}+{\bf{D}}^{4}_{x})(\tau_{1}.\tau_{1})=0
\label{2g}
\end{equation}
\begin{equation}
({\bf{D}}_{x}{\bf{D}}_{t}+{\bf{D}}^{4}_{x})(\tau_{2}.\tau_{2})=0
\label{2h}
\end{equation}
\begin{equation}
({\bf{D}}^{2}_{x})(\tau_{1}.\tau_{2})=0
\label{2i}
\end{equation}
The bilinearization for $N=1$ mKdV B, however, is not unique. It possesses
an alternate, but important set of bilinear forms
\begin{equation}
({\bf{D}}_{t}+{\bf{D}}^{3}_{x})(\tau_{1}.\tau_{2})=0
\label{2j}
\end{equation}
\begin{equation}
{\bf{D}}^{2}_{x}(\tau_{1}.\tau_{2})=0
\label{2k}
\end{equation}
The bilinear equations (\ref{2g},\ref{2h},\ref{2i}) while directly follow as
the reduction of the bilinear forms of the $N=1$ mKP equation
\begin{equation}
\partial_{t}U+D^{6}U-\frac{1}{2}D^{2}(U^{3})+12\partial_{y}^{2}D^{-2}U+6D^{2}U\partial_{y}D^{-2}U=0 
\label{2m}
\end{equation}
The equations (\ref{2j},\ref{2k}) become
useful to show its connection with bosonic mKdV equation. Both the equivalent
bilinear equations of the $N=1$ mKdV B equation do not involve the supersymmetric
Hirota operator leading different types of solitons solutions than those of
$N=1$ mKdV equation.

\setcounter{equation}{0}

\section{One Soliton Solutions}

The general structure of the $\tau$-functions for the one soliton solution 
both the $N=1$ mKdV equations may be written as
\begin{equation}
\tau_{1}=1+\alpha e^{\eta}
\label{3a}
\end{equation}
\begin{equation}
\tau_{2}=1+\beta e^{\eta}
\label{3b}
\end{equation}
$\alpha$ and $\beta$ in (\ref{3a},\ref {3b}) are nonzero constants and
\begin{equation}
\eta=kx+\omega t+\zeta \theta
\label{3c}
\end{equation}
where $k$ and $\omega$ are the bosonic parameters and $\zeta$ is the
Grassmann odd parameter. But the soliton solutions in their explicit forms
will be quite different.

Substituting (\ref{3a}) and (\ref{3b}) back into the Hirota equations of the $N=1$ 
mKdV equation (\ref{1i}) and (\ref{1j}), we find the non-trivial solutions 
provided
\begin{equation}
\beta=-\alpha.
\label{3d}
\end{equation}
The dispersion relation, however, follows from (\ref{1g}) and (\ref{1h}) as
\begin{equation}
\omega+k^{3}=0
\label{3e}
\end{equation}
which is identical to the bosonic mKdV equation and it is found that the 
fermionic parameter does not play a role at the one soliton solution level.
In obtaining the dispersion relation the following property of the super 
Hirota operator has been used.
\begin{equation}
{\bf{S}}{\bf{D}}^{n}\left(e^{\eta_{1}}.e^{\eta_{2}}\right)=\left(k_{1}-k_{2}
\right)^{n}\left[-\left(\zeta_{1}-\zeta_{2}\right)+\theta\left(k_{1}-k_{2}
\right)\right]e^{\eta_{1}+\eta_{2}}
\label{3f}
\end{equation}
The explicit form of the one soliton solution for the
superfield $\psi$ may be found by substituting the $\tau$-
functions in (\ref{1e}) and may be given as
\begin{equation}
\psi=\zeta {\mbox{cosech}}
(\phi + \gamma_0)-\theta k{\mbox{cosech}}
(\phi + \gamma_0)
\label{3g}
\end{equation}
where we have chosen $\phi = kx-k^3t$ and
$\alpha=-\beta=e^{\gamma_0}$, $\gamma_0$ being nonzero, real parameter. The
bosonic component of $\psi$ in (\ref{3g}) becomes indentical with the bosonic
mKdV solution as expected.

For the $N=1$ mKdV B equation (\ref{2}), by substituting the $\tau$ functions 
(\ref{3a},\ref{3b}) in the corresponding bilinear equations, lead to identical
constraint between the parameters (\ref{3d}) as well as the dispersion
relation (\ref{3e}). But the differerence between the soliton solutions of 
these two systems becomes apparent when we consider the two and higher 
soliton solutions. 

The one soliton solution of $N=1$ mKdV B equation in its component fields 
immediately follows from (\ref{2f}) as
\begin{equation}
\Psi=-k {\mbox{cosech}} (\phi + \gamma_0)-\theta (k\zeta )
{\mbox{cosh}}(\phi + \gamma_0){\mbox{cosech}}^{2}(\phi + \gamma_0)
\label{3h}
\end{equation}
where, $\phi = kx-k^3t$ and $\alpha=-\beta=e^{\gamma_0}$, $\gamma_0$ being 
nonzero, real parameter. Notice that the bosonic component in (\ref{3h}) 
becomes identical with that of (\ref{3g}), as expected, since both the 
equations reduce to the mKdV equation in the bosonic limit. But the fermion 
components are quite different.

\setcounter{equation}{0}

\section{$N$ Soliton Solution}

Existence of one soliton solution, in fact, does not ensure the exact 
integrability of the system. For that purpose, establishing the 
existence of soliton solutions upto three solitons becomes essential.
The general forms of the $\tau$ functions for $N$ soliton as before can be 
chosen to be same for the both the $N=1$ mKdV equations. The $\tau_1$ for 
the $N$ soliton may be written as
\begin{equation}
\tau_{1}=\sum_{\mu_{i}=0,1}exp\left(\sum_{i,j=1}^{N}\phi (i,j)\mu_{i}\mu_{j}
+\sum_{i=1}^{N}\mu_{i}(\eta_{i}+\log\alpha_{i})\right)
\;\;\;\;\;(i<j)
\label{4a}
\end{equation}
where $e^{\phi (i,j)}$ and $\alpha_{i}$ are the coefficients to be determined.
For convenience, we introduce $A_{ij}=e^{\phi (i,j)}$. We may write the form 
of the other $\tau$ function, namely $\tau_{2}$ by replacing $\alpha_{i}$ by
$\beta_{i}$ and $A_{ij}$ by $B_{ij}$.

In particular, for the two soliton solution $\tau$ functions acquire the form
\begin{equation}
\tau_{1}=1+\alpha_{1}e^{\eta_{1}}+\alpha_{2}e^{\eta_{2}}+\alpha_{1}\alpha_{2}
A_{12}e^{\eta_{1}+\eta_{2}}
\label{4b}
\end{equation}
and
\begin{equation}
\tau_{2}=1+\beta_{1}e^{\eta_{1}}+\beta_{2}e^{\eta_{2}}+\beta_{1}\beta_{2}
B_{12}e^{\eta_{1}+\eta_{2}}
\label{4c}
\end{equation}
where
\begin{equation}
\eta_{1}=k_{1}x+\omega_{1} t+\zeta_{1}\theta
\label{4d}
\end{equation}
and
\begin{equation}
\eta_{2}=k_{2}x+\omega_{2} t+\zeta_{2}\theta
\label{4e}
\end{equation}
In (\ref{4d},\ref{4e}) the parameters $k_1$, $k_2$, $\omega_1$ and $\omega_2$
are bosonic, while $\zeta_1$ and $\zeta_2$ are fermionic ones, as before.

For the $N=1$ mKdV equation, the bilinear equations (\ref{1i}) and (\ref{1j}) 
determine the following constraints on the parameters for nontrivial solution. 
The conditions that 
\begin{equation}
\beta_{i}=-\alpha_{i}
\label{4ee}
\end{equation}
and the dispersion relations
\begin{equation}
\omega_{i}+k^{3}_{i}=0
\label{4i}
\end{equation}    
for $i=1,2$ once again arise in the two soliton solution as in the case of 
one soliton. Moreover, these two bilinear equations yield the two
soliton interaction terms $A_{12}$ and $B_{12}$ as
\begin{equation}
A_{12}=B_{12}=\frac{(k_{1}-k_{2})^2}{(k_{1}+k_{2})^2}
\label{4f}
\end{equation}
In addition, (\ref{1j}) leads to a relation among the fermionic and bosonic
parameters, {\it viz.}
\begin{equation}
k_{1}\zeta_{2}=k_{2}\zeta_{1}
\label{4g}
\end{equation}                                                             
The condition (\ref{4g}) is necessary in order that (\ref{1g},\ref{1h}) are
consistently satisfied. It will be seen that the constraint on the parameters 
(\ref{4g}) is also crucial in demonstrating the existence of three soliton
solutions for the $N=1$ mKdV equation.                                         

For the $N=1$ mKdV B equation the bilinear equations (\ref{2g},\ref{2h},\ref{2i})
ensure the nontrivial two soliton solutions invoking the set of identical
results as in $N=1$ mKdV equation but for the last condition (\ref{4g}). In 
particular, the conditions on the parameter $\alpha_{i}$ and $\beta_{i}$ 
(\ref{4i}), the dispersion relations (\ref{4i}) and the interaction terms 
$A_{12}$ and $B_{12}$ (\ref{4f}) become identical in both the cases. However, 
in contrast to the $N=1$ mKdV, the relation among the fermionic and bosonic 
parameters does not arise for the $N=1$ mKdV B equation. This fact is  
reflected in the structure of the bilinear equations, which are expressed only in terms of the 
bosonic Hirota operator ${\bf D}$ and will be observed for all higher 
soliton solutions also.

The explicit forms of $\tau_{1}$ and $\tau_{2}$ for the three
soliton solution following (\ref{4a}) may be given as
\begin{eqnarray}
&&\tau_{1}=1+\alpha_{1}e^{\eta_{1}}+\alpha_{2}e^{\eta_{2}}+\alpha_{3}
e^{\eta_{3}}+\alpha_{1}\alpha_{2} A_{12}e^{\eta_{1}+\eta_{2}}+\alpha_{1}
\alpha_{3} A_{13}e^{\eta_{1}+\eta_{3}}\nonumber\\
&&+\alpha_{2}\alpha_{3} A_{23}e^{\eta_{2}+\eta_{3}}+\alpha_{1}\alpha_{2}
\alpha_{3} A_{12}A_{13}A_{23}e^{\eta_{1}+\eta_{2}+\eta_{3}}
\label{4k}
\end{eqnarray}
and
\begin{eqnarray}
&&\tau_{2}=1+\beta_{1}e^{\eta_{1}}+\beta_{2}e^{\eta_{2}}+\beta_{3}e^{\eta_{3}}
+\beta_{1}\beta_{2} B_{12}e^{\eta_{1}+\eta_{2}}+\beta_{1}\beta_{3}
B_{13}e^{\eta_{1}+\eta_{3}}\nonumber\\
&&+\beta_{2}\beta_{3} B_{23}e^{\eta_{2}+\eta_{3}}+\beta_{1}\beta_{2}\beta_{3}
B_{12}B_{13}B_{23}e^{\eta_{1}+\eta_{2}+\eta_{3}}
\label{4l}
\end{eqnarray}
where
\begin{equation}
\eta_i= k_{ix}x+\omega_i t+\zeta_i \theta \quad(i=1,2,3)
\label{4m}
\end{equation}
Notice that (\ref{4k},\ref{4l}) do not contain any new unknown parameter; it 
is expressed in terms of the parameters of the two soliton solutions only. 
Three soliton solutions thus verifies the consistency of the parameters 
determined by one and two soliton solutions. Substitution of the three 
soliton solutions in the bilinear equations gives rise to a set of nontrivial 
relations among the parameters, which determine the consistency of the 
solutions. In the three soliton solution we find the conditions 
$\beta_{i}=-\alpha_{i}$ and the dispersion relations $\omega_{i}+k^{3}_{i}=0$ 
$(i=1,2,3)$ for all the three solitons for both the $N=1$ mKdV and the $N=1$ 
mKdV B equations. The interaction terms for both the equations become
\begin{equation}
A_{ij}=B_{ij}=\frac{(k_{i}-k_{j})^2}{(k_{i}+k_{j})^2} \quad (i,j=1,2,3;\quad
i\neq j)
\label{4n}
\end{equation}
The $N=1$ mKdV equation however admits additional constraint as in the two 
soliton solution on the fermionic parameters:
\begin{equation}
k_{i}\zeta_{j}=k_{j}\zeta_{i} \quad (i,j=1,2,3;\quad i\neq j)
\label{4o}
\end{equation}
which are not found for the $N=1$ mKdV B equation. The constraints (\ref{4o}) 
are essential to ensure the coefficients of the terms $e^{\eta_1+\eta_2}$,
$e^{\eta_1+\eta_3}$, $e^{\eta_2+\eta_3}$, $e^{2\eta_1+\eta_2+\eta_3}$,  
$e^{\eta_1+2\eta_2+\eta_3}$ and $e^{\eta_1+\eta_2+2\eta_3}$ to vanish. 
Apart from determing the unknown parameters, three soliton 
solutions provide a nontrivial relation among the parameters. This identity 
follows from the coefficient of the term $e^{\eta_1+\eta_2+\eta_3}$ to be 
zero and is given by 
\begin{eqnarray}
A_{12}A_{13}A_{23}[(\omega_1+\omega_2+\omega_3)+(k_1+k_2+k_3)^{3}]
[-(\zeta_1+\zeta_2+\zeta_3)+\theta (k_1+k_2+k_3)]&& \nonumber\\
+A_{23}[(\omega_1-\omega_2-\omega_3)+(k_1-k_2-k_3)^{3}]
[-(\zeta_1-\zeta_2-\zeta_3)+\theta (k_1-k_2-k_3)]&& \nonumber\\
+A_{13}[(\omega_2-\omega_1-\omega_3)+(k_2-k_1-k_3)^{3}]
[-(\zeta_2-\zeta_1-\zeta_3)+\theta (k_2-k_1-k_3)]&& \nonumber\\
+A_{12}[(\omega_3-\omega_1-\omega_2)+(k_3-k_1-k_2)^{3}]
[-(\zeta_3-\zeta_1-\zeta_2)+\theta (k_3-k_1-k_2)]=0&& \nonumber\\ 
\label{4p}
\end{eqnarray}
and interestingly is satisfied by using the dispersion relations and 
the parameters in (\ref{4o}). In fact, (\ref{4p}) makes the three soliton 
solution nontrivial.

Existence of $N$ soliton solutions may be shown following the same procedure 
as the three soliton solutions for both the systems. Apart from verifying the
unknown parameters, determined by one and two soliton solutions, $N$ soliton
solutions provide a set of identities to be satisfied involving the 
parameters of one and two soliton solutions. However, for odd $N$ is found 
that terms of even degree become trivial as in the three soliton solution 
case. By the degree of a term we mean the total number of $\eta_i$ present 
in the exponent of a term. The most nontrivial identity for odd $N$ follows
from the coefficient of the term involving all the $\eta_i$ appearing once 
only. For even soliton solutions, on the other hand, the terms of odd degree 
become trivial.

\section{Conclusion}

Two important dynamical systems, $N=1$ mKdV and $N=1$ mKdV B have been shown
to possess soliton solutions. The soliton solutions of these systems have 
been obtained following the super extension of the Hirota bilinear formalism.
It is found that the bilinear forms of $N=1$ mKdV system involve super Hirota 
derivative, whereas the bilinear forms of $N=1$ mKdV B system can be 
expressed in the terms of bosonic Hirota derivative only. This difference of
the two systems of motion is also reflected in their soliton solutions from 
two soliton solutions onwards. While the former equation of motion gives 
rise to a set of conditions on the fermionic parameters for the nontrivial 
solutions to exist, the latter does not. The fermion components of the soliton 
solutions for these two systems become quite different. The bosonic 
components, however,  become identical for both the cases. This is in 
conformation with the fact that both the systems reduce the same mKdV 
equation in the bosonic limit, although they have different $N=1$ 
supersymmetric versions. These two equations also follow from two separate 
$N=2$ dynamical systems as an $N=1$ reduction. 

\vspace{1 cm}

{\it SG would like to thank DST, Govt. of India for financial support under
the project no. 100/(IFD)/2066/2000-2002.}

\end{document}